\documentclass[journal,twoside,web]{ieeecolor}
\usepackage{generic}
\usepackage{cite}
\usepackage{amsmath,amssymb,amsfonts}
\usepackage{algorithmic}
\usepackage{graphicx}
\usepackage{textcomp}
\usepackage{pdfpages}
\usepackage{multirow}
\usepackage{easytable}
\usepackage{hyperref}

\usepackage[flushleft]{threeparttable} 
\usepackage{booktabs,caption}

\def\BibTeX{{\rm B\kern-.05em{\sc i\kern-.025em b}\kern-.08em
    T\kern-.1667em\lower.7ex\hbox{E}\kern-.125emX}}
\begin{document}

\title{Uncertainty-Aware Semi-Supervised Learning for Prostate MRI Zonal Segmentation}
\author{Matin Hosseinzadeh, Anindo Saha, Joeran Bosma, Henkjan Huisman
    \thanks{This research is supported by Siemens Healthineers (CID: C00225450) and the PIONEER H2020 European project.}
    \thanks{M. Hosseinzadeh, A. Saha, J. Bosma, H. Huisman are with the Diagnostic Image Analysis Group of the Department of Imaging, Radboudumc, 6525GA Nijmegen, The Netherlands (e-mail: matin.hosseinzadeh@radboudumc.nl; anindya.shaha@radboudumc.nl; joeran.bosma@radboudumc.nl; henkjan.huisman@radboudumc.nl)}
    }

\maketitle

\begin{abstract}
Quality of deep convolutional neural network predictions strongly depends on the size of the training dataset and the quality of the annotations. Creating annotations, especially for 3D medical image segmentation, is time-consuming and requires expert knowledge. We propose a novel semi-supervised learning (SSL) approach that requires only a relatively small number of annotations while being able to use the remaining unlabeled data to improve model performance. Our method uses a pseudo-labeling technique that employs recent deep learning uncertainty estimation models. By using the estimated uncertainty, we were able to rank pseudo-labels and automatically select the best pseudo-annotations generated by the supervised model. We applied this to prostate zonal segmentation in T2-weighted MRI scans. Our proposed model outperformed the semi-supervised model in experiments with the ProstateX dataset and an external test set, by leveraging only a subset of unlabeled data rather than the full collection of 4953 cases, our proposed model demonstrated improved performance. The segmentation dice similarity coefficient in the transition zone and peripheral zone increased from 0.835 and 0.727 to 0.852 and 0.751, respectively, for fully supervised model and the uncertainty-aware semi-supervised learning model (USSL). Our USSL model demonstrates the potential to allow deep learning models to be trained on large datasets without requiring full annotation. Our code is available at https://github.com/DIAGNijmegen/prostateMR-USSL.
\end{abstract} 

\begin{IEEEkeywords}
deep learning, prostate segmentation, semi-supervised, uncertainty
\end{IEEEkeywords}

\section{Introduction}
\label{sec:introduction}
\IEEEPARstart{M}EDICAL image segmentation plays an important role in computer-assisted diagnosis and surgical planning \cite{3DUNet}. Deep learning-based approaches have achieved great success in supervised learning tasks where labeled data is abundant \cite{litjens2017survey}. However, acquiring a large amount of accurately annotated data is time-consuming, labor-intensive, and often requires expert knowledge. Without expert annotation, the current supervised deep learning models cannot learn from extensive medical imaging data. In this paper, we aim to explore a semi-supervised model to reduce the labor of large-scale 3D volumetric annotation, by effectively leveraging the unlabeled data.

Semi-Supervised Learning (SSL)  is a type of learning that combines supervised and unsupervised approaches to improve performance on small datasets by leveraging large quantities of unlabeled data \cite{verma2019interpolation}. 

In the field of medical image analysis, SSL methods have been widely studied as a way to reduce the labor-intensive process of manual annotation\cite{litjens2017survey}. These methods can be broadly grouped into two categories: consistency-based approaches, which use augmentation and perturbation to encourage the model to produce consistent predictions on different versions of the same input data, and self-learning approaches, which use the model's own predictions on unlabeled data to generate pseudo-labels and train the model in a self-taught manner. In this paper, we propose a novel SSL algorithm called uncertainty-aware semi-supervised learning (USSL), which uses predictive uncertainty estimation to select the most accurate pseudo-labels and improve the segmentation model.

A reliable estimation of the predictive uncertainty of deep learning in medical imaging is vital in order to effectively use the system. Overconfident incorrect predictions may lead to misdiagnoses or sub-optimal treatment, hence proper uncertainty estimation is crucial for practical applications in medicine. In most cases, it is difficult to evaluate the quality of predictive uncertainties, since the \lq{ground truth}\rq of uncertainty estimation is usually unavailable \cite{lakshminarayanan2017simple}. We can ask human readers how confident they are of a particular prediction, but deep learning typically relies on softmax outputs rather than binarized segmentations to provide a measure of uncertainty. However, in practice, deep learning models can still be prone to overfitting the training data. To obtain a more reliable estimate of uncertainty, a deep learning model that generates a distribution of possible outcomes can be used \cite{filos2019systematic}. Most deep learning models are often deterministic functions, and as a result, are operating in a very different setting to probabilistic models which can also learn uncertainty information \cite{gal2016uncertainty}. 

Deep learning algorithms are being developed to incorporate uncertainty, such as the use of Bayesian neural networks and Monte Carlo Dropout methods. \cite{caldeira2020deeply}. Monte Carlo Dropout (MC-Dropout) is a technique used to avoid over-fitting in neural networks \cite{gal2016uncertainty}. Gal \textit{et al.} proposed MC-Dropout to estimate predictive uncertainty by using dropout at inference time. They showed that optimizing any neural network with the standard dropout regularization and L2-regularization is equivalent to a form of variational inference in a probabilistic interpretation of the model \cite{gal2016uncertainty}. Another technique for uncertainty estimation is Deep Ensembles \cite{lakshminarayanan2017simple}. This method averages out the predictions of several deterministic models. It is simple to implement and can be parallelized easily. 

Automated segmentation of prostate transition zone (TZ) and peripheral zone (PZ) from T2-weighted MRI scans plays an essential role in clinical diagnosis, treatment planning, and improving automated CAD tools \cite{hosseinzadeh2021deep,SAHA2021102155}. However, prostate zonal segmentation is challenging because it is a complex organ with varying size, shape and appearance, fuzzy borders, and poor image contrast at the boundary \cite{montagne2021challenge}. Several studies reported difficulty and inter-observer variability of prostate zonal segmentation \cite{montagne2021challenge,becker2019variability}.

In this paper, we propose a novel USSL algorithm to address the above issues and apply this to prostate zonal segmentation. 
We hypothesize that combining an uncertainty-aware segmentation method and an SSL-based technique can enable the deep learning model to learn from an extensive dataset of unlabeled scans by exploiting the uncertainty information. We propose to use a probabilistic fully convolutional neural network to model the ambiguity in the labels and original images. Our proposed framework is applied for prostate zonal segmentation in T2W images.
The main contributions of this paper are summarized below: 
\begin{itemize}
\item We propose an uncertainty estimation method using a probabilistic model that outputs uncertainty for each predicted zone.
\item We show that our predictive uncertainty method yields reliable calibration of model uncertainty that correlates inversely with the segmentation quality metrics and does not require a ground-truth label. 
\item We propose USSL to improve the existing SSL model by leveraging the estimated uncertainty for unlabeled data.
\item We validate our USSL model on the prostate zonal segmentation task, using two different test sets, and show that our model obtains significant improvements over SL and SSL methods.
\end{itemize}

\section{Related Works}
In this section, we provide a brief review of recent research on semi-supervised learning and predictive uncertainty estimation, and their potential applications to prostate segmentation.

\subsection{Prostate zone segmentation}
Many classical methods have been proposed for automated prostate zonal segmentation including atlas-based segmentation \cite{litjens2012pattern}, active appearance models \cite{toth2013simultaneous} and pattern recognition approaches \cite{makni2011zonal}.  Currently, CNNs are the most popular and cutting-edge approach for prostate segmentation \cite{aldoj2020automatic,zabihollahy2019automated,cuocolo2021deep}. Most CNN-based models employ variants or extensions of the 2D or 3D U-Net models \cite{ronneberger2015u,3DUNet}. Aldoj \textit{et al.} \cite{aldoj2020automatic} used a Dense-2 U-net to segment PZ and TZ on axial T2-weighted image using coarsely and fine annotated segmentation masks, to study the impact of ground truth variability on segmentation. In another research two parallel U-net models were used to segment the prostate and its zones on T2w and ADC scans \cite{zabihollahy2019automated}. Cuocolo \textit{et al.} \cite{cuocolo2021deep} compared efficient neural network (ENet) and efficient residual factorized ConvNet (ERFNet) to segment prostate zones and reported Dice scores of $87\%\pm5\%$ and $71\%\pm8\%$ for TZ and PZ, respectively. These analyses were performed using a small number of testing images because annotating 3D prostate scans typically takes a considerable amount of time and effort.

\subsection{Semi-supervised medical image segmentation}
Semi-supervised learning (SSL) is a naturally occurring scenario in medical imaging. In segmentation methods, an expert reader might label only a part of the data, leaving many samples unlabeled \cite{cheplygina2019not}. Recent semi-supervised deep learning methods in the medical image analysis domain mostly use self-training or co-training approaches \cite{cheplygina2019not}.

Self-training is a popular approach for SSL in medical imaging that uses label propagation \cite{cheplygina2019not}. In self-training, a model is trained using labeled data and then used to create pseudo-labels for the rest of the data. Subsequently, these samples, or a subset of them, are used as part of the training set \cite{cheplygina2019not}. For segmentation, self-training is popular for pixel/voxel label propagation. It is used in the brain \cite{meier2014patient,wang20144d}, retina \cite{gu2017semi}, heart \cite{bai2017semi} and several other applications. In addition to self-training several papers choose an active learning approach, where experts verify some of the labels \cite{parag2014small,su2015interactive}. 

Overall, recent works have shown that SSL can improve performance in medical image segmentation tasks, especially when labeled data is limited. However, these methods are still limited by the quality of the pseudo-labels generated by the model.

\subsection{Uncertainty estimation}
Recent trends have shown an increased interest in uncertainty estimation and confidence measurement with deep neural networks. For uncertainty estimation, Deep Ensembles \cite{lakshminarayanan2017simple}, Monte-Carlo Dropout \cite{gal2016dropout}, and stochastic variational Bayesian inference \cite{zhang2018advances} have been proposed. A recent paper by Meyer \textit{et al.} \cite{meyer2021uncertainty} proposed an uncertainty-aware temporal self-learning method. In their model, they use temporal ensembling and uncertainty-guided self-learning to segment prostate zones. Liu \textit{et al.} \cite{liu2020exploring} used a Bayesian deep learning network to model the long-range spatial dependence between PZ and TZ in prostate MRI. Mehrtash \textit{et al.} \cite{mehrtash2020confidence} proposed model ensembling for confidence calibration of the FCNs trained with batch normalization and used a calibrated FCN to measure prostate whole gland segmentation quality and detect out-of-distribution test examples.

\section{Methods}
In this section, we detail our proposed method, including the datasets, network architecture, key components, and evaluation metrics. We introduce the uncertainty-aware semi-supervised learning (USSL) approach and explain how it utilizes uncertainty estimation and a semi-supervised framework to selectively expand the training dataset size and enhance the accuracy of segmentation.

\subsection{Data and pre-processing} 
We used three different datasets to train and test our model. To train our supervised models, we used 200 prostate T2W MRI scans from the publicly-available ProstateX dataset \cite{armato2018prostatex}, paired with voxel-level delineations of the whole-gland (WG), central zone + anterior stroma + transition zone (TZ), peripheral zone (PZ) annotated by experienced radiologists \cite{cuocolo2021quality}. We used this dataset also for testing using 5-fold cross-validation (see section \ref{sec: Implementation details} for more details).

To train our semi-supervised model we used a large internal cohort of 4953 unlabeled MRI scans from 4357 patients. This consecutive, regular clinical mpMRI dataset was acquired from 2014 to 2020 at Radboudumc, Nijmegen, The Netherlands.
In addition, the trained models are further evaluated on 111 scans from an external prostate mpMRI dataset acquired at St. Olavs Hospital, Trondheim University Hospital, Trondheim, Norway \cite{kruger2021multiparametric} to test the robustness of the model. All scans in this dataset were annotated by a radiology resident under the supervision of an expert radiologist (At least 10 years of experience in prostate MRI). We will refer to this dataset as the \textit{external} test set.

All scans were acquired on 3T MR scanners (MAGNETOM Trio and Skyra, Siemens Healthineers) using a turbo spin-echo sequence with 0.3-0.5 mm in-plane resolution and 3.6 mm slice thickness. To ensure consistency, all images were resampled to 0.5$\times$0.5$\times$3.6 mm$^3$ resolution, and then cropped at the center to create images of size 160$\times$160$\times$20 voxels. To normalize the image intensities, we used z-score normalization to ensure that all images have a mean of 0 and a standard deviation of 1. \cite{saha2021anatomical}.

\subsection{Problem formulation} 
We formalize the problem of semi-supervised 3D image segmentation as follows.
Given labeled dataset $S_{l}=\left\{\left(x_{1}^{l}, y_{1}^{l}\right), \ldots,\left(x_{k}^{l}, y_{k}^{l}\right)\right\}$, which contains $k$ labeled examples, each example comprised of an input image $x_{i}^{l} \in \mathbb{R}^{H\times W\times D}$ and its corresponding pixel-level label $y_{i}^{l} \in \mathbb{R}^{H\times W\times D\times C}$, where $C$ is the number of classes and $H\times W\times D$ is spatial dimension. In a semi-supervised setting, we also have an unlabeled set of images $\left\{x_{1}^{u}, \ldots, x_{n}^{u}\right\}$ typically $n$ unlabeled images, with $n \gg k$. Using a trained model we can create pseudo-labels for each input $x_{i}^{u}$ as $\widehat{y}_{i}^{u}$ forming $S_{u}=\left\{\left(x_{1}^{u}, \widehat{y}_{1}^{u}\right), \ldots,\left(x_{n}^{u}, \widehat{y}_{n}^{u}\right)\right\}$.
The purpose of semi-supervised segmentation is to train a segmentation model $f$ with parameters $\theta$ and with $S=S_{l} \cup S_{u}$, to map each pixel of an input image to its label.

\subsection{Architectural details}  \label{sec:Architectural}
\begin{figure*}[!t]
    \centerline{\includegraphics[width=0.85\textwidth]{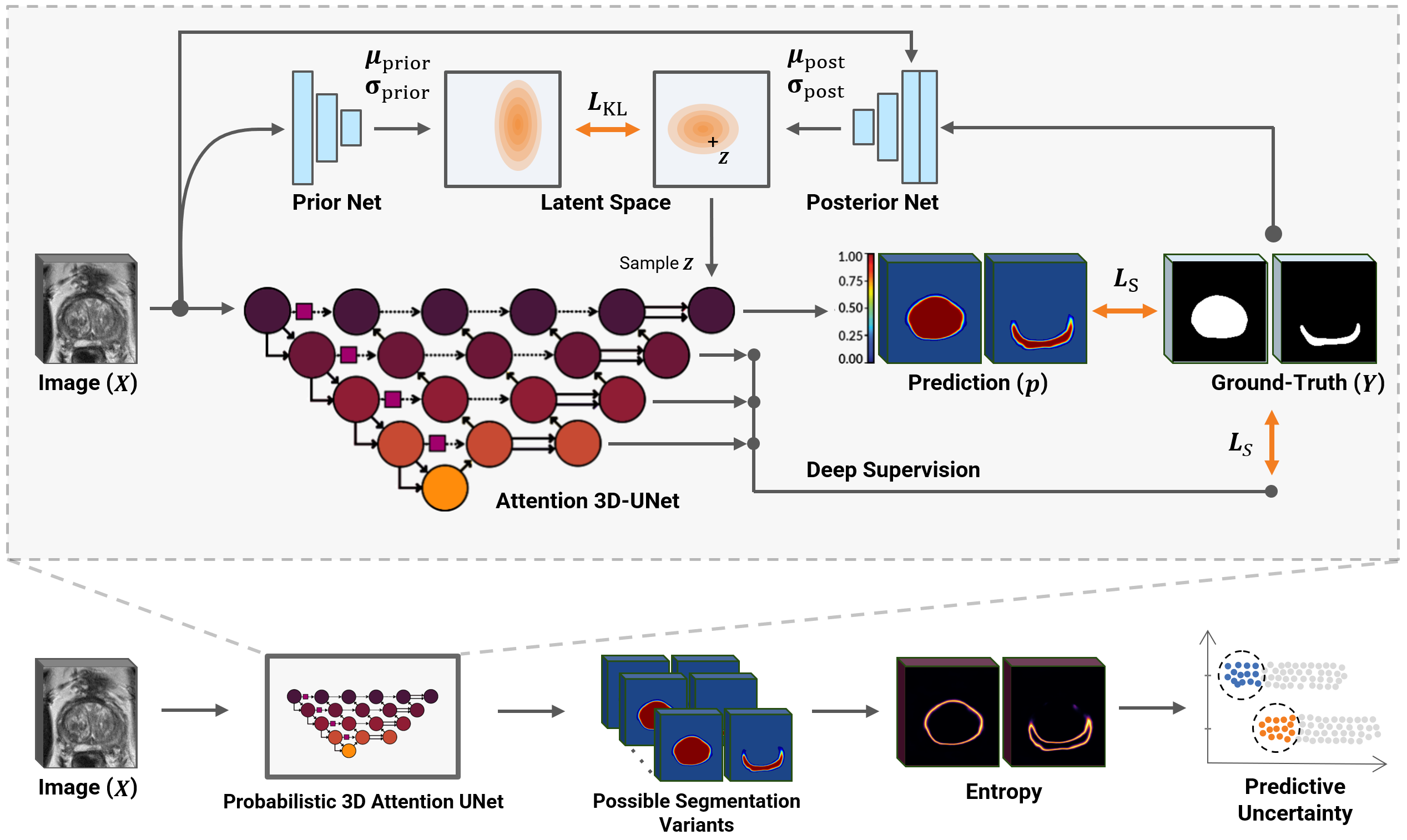}}
    \caption{Schematic illustration of our probabilistic attention UNet (PA-UNet) model for segmentation.}
    \label{fig:PAUNET_framework}
\end{figure*}

In this section, we introduce our probabilistic segmentation framework, which has the capability to generate multiple segmentation hypotheses for a given input image. We illustrate the overall architecture in Fig. \ref{fig:PAUNET_framework}. We used a probabilistic adaptation (as specified by \cite{ProbUNet}) of the deep attentive 3D U-Net (PA-UNet) \cite{SAHA2021102155}, that was developed and validated specifically for prostate MRI \cite{saha2021anatomical}. We introduce deep supervision in PA-UNet to learn robust features even in the early layers. This method adds a companion objective function at the earlier stages of the UNet's encoder in addition to the overall objective function at the output layer. Our model employs conditional variational autoencoders (CVAE) adopted from \cite{ProbUNet} that are capable of modeling complex distributions and producing numerous plausible segmentations by encoding them to a low-dimensional latent space \cite{ProbUNet} and drawing a random sample to predict a segmentation mask. During inference, a sample $z$ from posterior distribution $J$ combines with the activation map of Attention UNet. Monte-Carlo Dropout was also added to capture both \textit{epistemic} and \textit{aleatoric} uncertainty during inference (as it is recommended by \cite{MCProb}). 
The source code of the model is publicly available\footnote[1]{https://github.com/DIAGNijmegen/prostateMR-USSL}.

\noindent \textbf{Losses and Objectives.} 
The network is trained with three constraints to simultaneously learn segmentation and its variations:

\begin{equation}
    \mathcal{L}_{\text {total}}=\lambda_{1} \mathcal{L}_{{S}}+\lambda_{2} \mathcal{L}_{{DS}}+\lambda_{3} \mathcal{L}_{{KD}}
    \label{eq_loss}
\end{equation}

\noindent where $\lambda_{1}$, $\lambda_{2}$ and $\lambda_{3}$ are weighting factors for segmentation loss ($\mathcal{L}_{{S}}$), deep supervision loss ($\mathcal{L}_{{DS}}$) and CVAE loss ($\mathcal{L}_{{KD}}$). We determine these hyper-parameters by optimizing on the validation set. In Eq. \eqref{eq_loss}, $\mathcal{L}_{{S}}$ cross-entropy loss penalizes pixel-wise differences between the softmax output ($\hat Y$) and ground-truth ($Y$) as defined by:

\begin{equation}
  \begin{aligned}
  \operatorname{\mathcal{L}_{{S}}}\left(Y, \hat Y\right)=&-\beta Y \log \hat Y -(1-\beta)\left(1-Y\right) \log \left(1-\hat Y\right)
  \end{aligned}
\end{equation}

\noindent in addition, as a standard practice for VAEs, we use Kullback-Leibler divergence loss ($\mathcal{L}_{\mathrm{KL}}(Q \| J)$) to penalize variance between the posterior distribution $Q$ and the prior distribution $J$ by maximizing the so-called evidence lower bound (ELBO). By training with this KL loss, we ‘pull’ the posterior distribution and the prior distribution toward each other.

\noindent \textbf{Predictive Uncertainty Estimation.} By using the probabilistic model explained in this section, we run \(T\) stochastic forward passes using the trained probabilistic model for each 3D input volume. In each forward pass, a random sample from latent space is injected into the segmentation model to produce a segmentation variation. As a result, we acquire a collection of softmax probabilities for each voxel in the input $\left\{\mathbf{p}_{t}\right\}_{t=1}^{T}$. The uncertainty of this probability vector $\mathbf{p}$ can then be summarized using the entropy of the probability vector, $E$, for each class \(c\) \cite{kendall2017uncertainties}:

\begin{equation}
\mu_{c}=\frac{1}{T} \sum_{t} \mathbf{p}_{t}^{c} 
\label{eq1}
\end{equation}

\begin{equation}
E_{c}= \mu_{c} \log \mu_{c}
\label{eq2}
\end{equation}

\noindent where ${p}_{t}^{c}$ is the probability of the ${c}$-th class in the ${t}$-th time prediction and $E_c$ is the predicted entropy for class $c$.

Motivated by the uncertainty estimation in \cite{mehrtash2020confidence,yu2019uncertainty} we propose to use the mean of entropy inside the predicted zone as a metric for assessing the quality of segmentation. Entropy captures the average amount of information contained in the predictive distribution \cite{gal2016uncertainty}.
Entropy is high when the input is ambiguous, indicating high aleatoric uncertainty. Alternatively, entropy can also be high when a probabilistic model has many possible explanations for the input, indicating high epistemic uncertainty \cite{filos2019systematic,kendall2017uncertainties}. 
In our case, we can estimate predictive entropy by collecting the probability vectors from ${T}$ stochastic forward passes. To obtain volumetric uncertainty for each class $c$, we summarize $\left\{{p}_{t}\right\}_{t=1}^{T}$ by computing predictive entropy separately for each class.

\subsection{Uncertainty-aware semi-supervised segmentation} 
\label{Uncertainty-aware Semi-supervised Segmentation}
Many successful semi-supervised learning approaches build upon generating pseudo-labels for unlabeled data using a supervised model trained on the labeled data. Typically, these approaches learn representations by extending the training data and improving the generalization of the model. However, many of generated pseudo-labels are incorrect, leading to noisy training data and unsatisfactory generalization. This is especially concerning when the task is complex and the supervised model is unable to achieve a high level of performance. We discovered that selecting predictions with a low level of uncertainty using the proposed method in \ref{sec:Architectural}, decreases the noise in the training data and improves generalization.

\begin{figure}[!t]
\centerline{\includegraphics[width=\columnwidth]{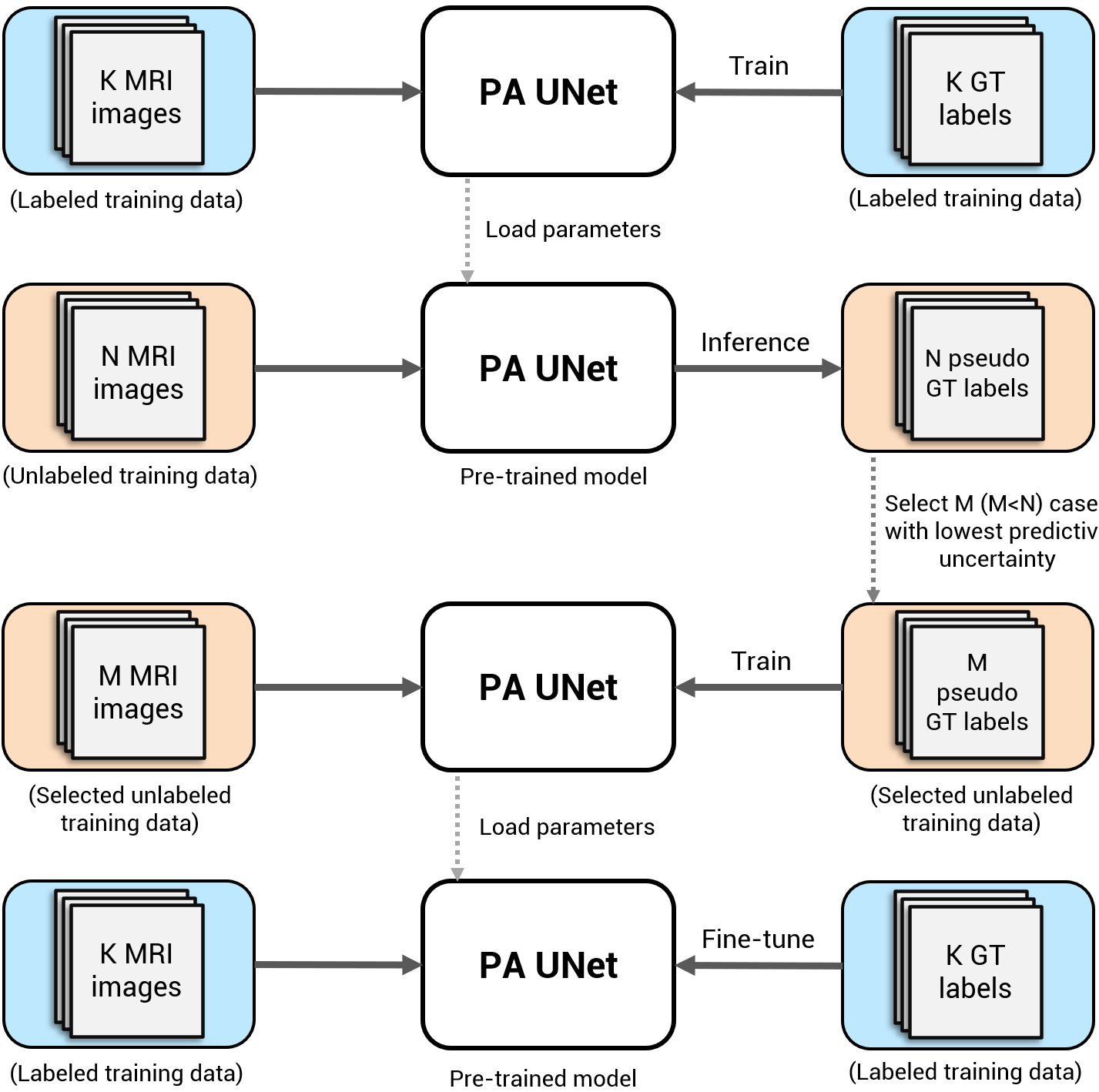}}
\caption{Overview of the proposed Uncertainty-Aware Semi-supervised Learning (USSL) framework. The training process is stepwise and starts with the top model in the figure and progresses downwards to the bottom model.}
\label{fig:overview_USSL}
\end{figure}

We present an uncertainty-aware pseudo-label selection method by using the most confident network outputs and use a less noisy subset of pseudo-labels for training the second model. In our semi-supervised approach, the probabilistic model is trained on labeled data and used to predict pseudo-labels for all the unlabeled data. We use the average of ${T=20}$ network predictions for each input image to obtain pseudo labels for unlabeled data. Using the estimated uncertainty $E_{c}$ we can estimate the predicted segmentation quality without having the ground-truth label for each class $c$. In our setting, we filter out the cases with the highest predictive uncertainty in the unlabeled set and keep the most certain cases $max(E_c^n) \leq \tau_{c}$. A second model is then trained on the cases with certain pseudo-labels. An overview of our USSL framework is shown in Fig. \ref{fig:overview_USSL}

\subsection{Evaluation metrics}
The quality of a 3D segmentation can be assessed using various metrics such as Dice similarity coefficient (DSC), Hausdorff Distance, Volume/Mass Error, Relative Absolute Volume Difference (RAVD),  among others \cite{fournel2021medical}. In this study, we assess the quality of our segmentation by using the DSC and RAVD, which are commonly used metrics in the medical imaging community. We report the DSC for each of the prostate zones (PZ and TZ) in different areas of the prostate (the apex, mid-gland, and base) to give a detailed understanding of the segmentation performance:
\begin{equation}
DSC=\frac{2|A \cap B|}{|A|+|B|}
\end{equation}
where A is the predicted 3D segmentation and B is the ground-truth manual segmentation. The DSC gives a measurement value in the range of [0,1], where the minimum and maximum values correspond to no overlap and perfect match, respectively. For RAVD, a lower value indicates a better agreement between the two volumes, as it means that the difference between the two volumes is smaller.

We can estimate the level of confidence in the predicted segmentation by the uncertainty measure ($E_c$). If high model uncertainty correlates with erroneous predictions, this information could be leveraged to mimic clinical quality control workflow. We can evaluate the uncertainty metric by ranking the prediction based on estimated uncertainty and comparing that to the actual DSC. We validate the uncertainty measure in section \ref{sec:Architectural} by correlating the predicted entropy with the segmentation quality (DSC). 

To determine whether the results of one model are significantly different from another model, we performed a statistical significance test using paired t-tests. This test compares the means of the performance metrics for the two models, and a p-value less than 0.05 indicates that the difference between the means is statistically significant.

\subsection{Implementation details}  \label{sec: Implementation details}
In our implementation, we used a dropout rate of 0.3. During the training, we used Cosine annealing learning rate \cite{CosineAnnLRWR} (oscillating between 10$^{\textnormal{-}4}$ to 10$^{\textnormal{-}7}$) and \textit{AMSGrad} optimizer \cite{AMSBound}. Gaussian noise (standard deviation 0-0.5), rotation (max. $\pm$7.5\textdegree), horizontal flip, translation (0-15\% horizontal/vertical shifts) and scaling (0-15\%) centered along the axial plane were used as data augmentation techniques. We train all models with batch size 1 and for 150 epochs. All experiments were performed on a single NVIDIA GTX 2080 Ti GPU, and implemented using TensorFlow 2.
\label{Implementation Details}
\subsubsection{Pseudo label generation} We generated pseudo-labels by running the probabilistic model for $T=20$ stochastic forward passes and averaging the outputs.
\subsubsection{Class imbalance} We compensate for the class imbalance between different classes by using the weighted cross-entropy as the cost function, attributing more weight to the classes with smaller regions. We used 0.05, 0.3, and 0.65 factors to re-weight the weighted cross-entropy for WG, TZ, and PZ, respectively.

\subsection{Comparison with state-of-the-art methods} 
\label{Comparison with state-of-the-art methods}
We have compared our proposed uncertainty-aware semi-supervised learning approach against state-of-the-art segmentation methods in terms of DSC. There are several factors that change between the USSL and deterministic fully supervised training. In order to evaluate the importance of each factor, we progressively move from the 3D-UNet setting to the USSL setting.

\subsubsection{3D-UNet} We used the 3D-UNet model as a baseline \cite{3DUNet} which extends the original 2D-UNet architecture to 3D \cite{ronneberger2015u}. This model is a common performance baseline for image segmentation. 
\subsubsection{Attention 3D-UNet} This model is based on attention 3D U-Net \cite{SAHA2021102155}, trained without dropout, using exactly the same set-up and hyper-parameters as the probabilistic model.

\section{Results}
\label{results}
\begin{figure*}[!t]
    \centerline{\includegraphics[width=0.85\textwidth]{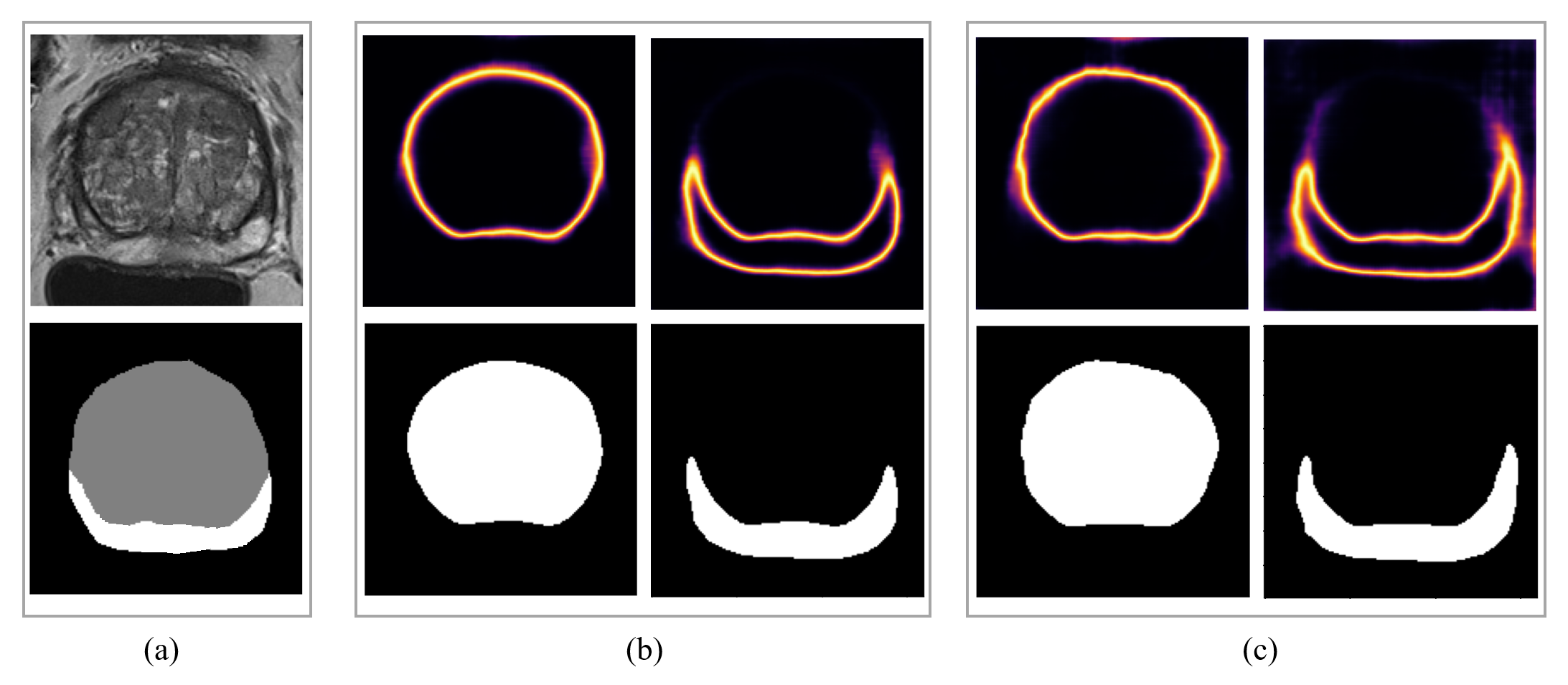}}
    \caption{Visual comparison of prostate zonal segmentation results for the same image with an without USSL method. (a) T2-weighted MRI image with ground truth segmentation of transition zone (TZ) and peripheral zone (PZ). (b) Probabilistic U-Net (PUNet) with uncertainty-aware semi-supervised learning (USSL) method, showing both the segmentation mask and the corresponding uncertainty map for each zone. (c) PUNet trained in a supervised manner, also displaying the segmentation mask and uncertainty map for each zone. The USSL method demonstrates improved segmentation performance and a reduction in uncertainty, particularly in more difficult regions, resulting in a better representation of the uncertainty in the predicted zones.}
    \label{fig:visual_results}
\end{figure*}

\begin{figure}[!t]
    \centerline{\includegraphics[width=\columnwidth]{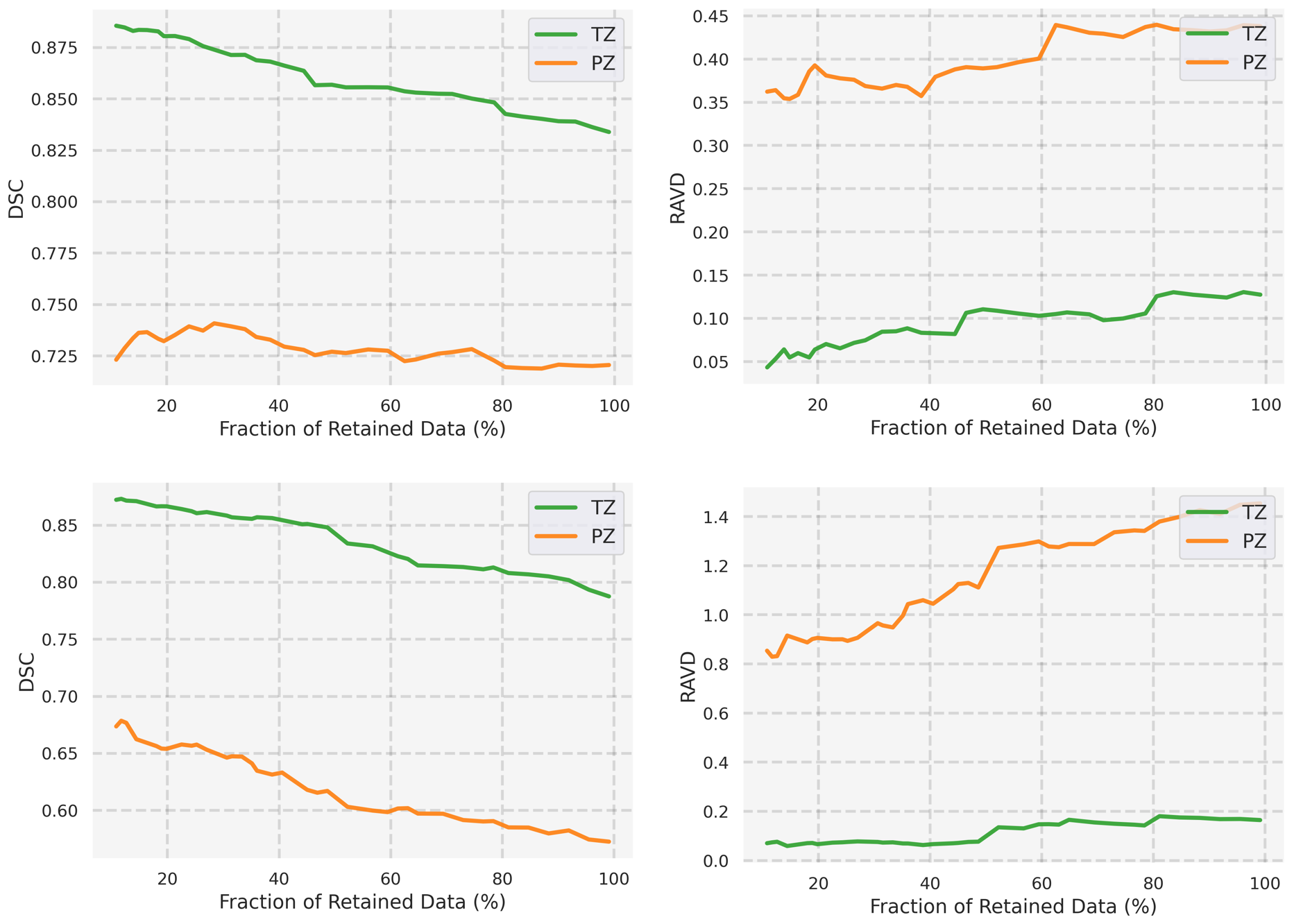}}
    \caption{Dice Similarity Coefficient (DSC) and Relative Absolute Volume Difference (RAVD) vs Fraction of the retained data (\%) \textbf{(\textit{Top})} ProstateX test set \textbf{(\textit{Bottom})} the external test set}
    \label{fig:dsc_trend}
\end{figure}

\begin{figure}[!t]
    \centerline{\includegraphics[width=\columnwidth]{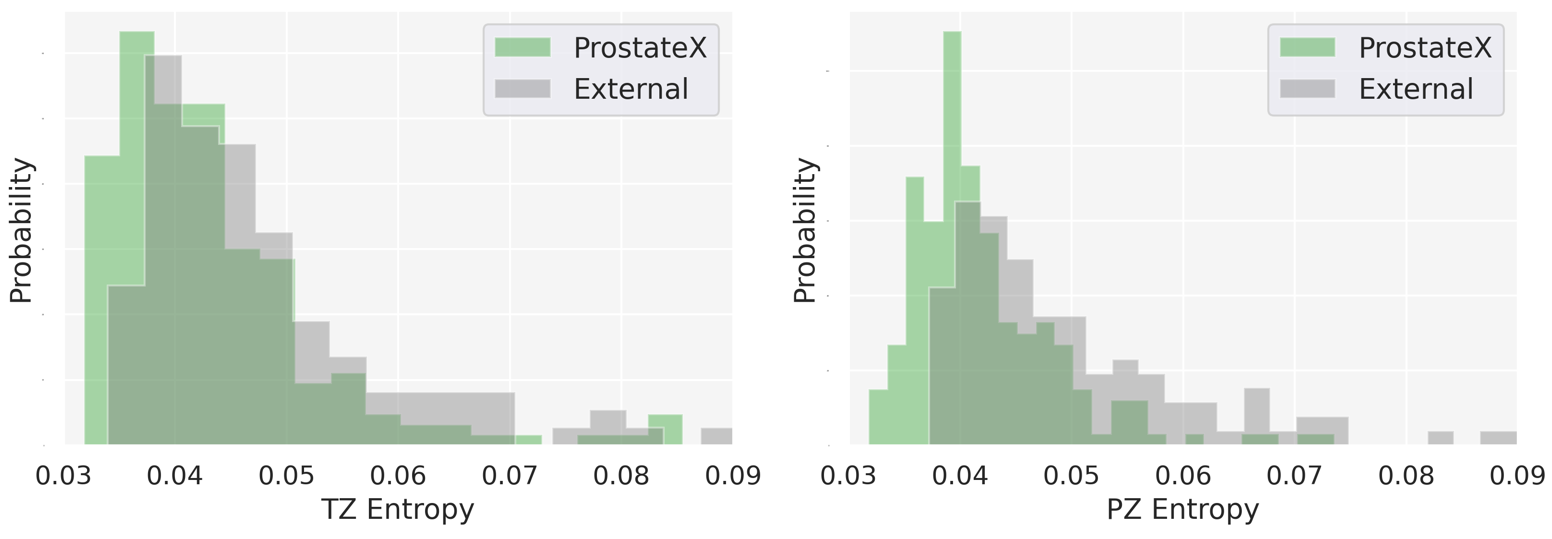}}
    \caption{Comparison of uncertainty histograms between two in-distribution (ProstateX) and out-distribution (External) test sets for TZ (\textit{left}) and PZ (\textit{right}).}
    \label{fig:hist}
\end{figure}

\begin{figure}[!t]
\centerline{\includegraphics[width=\columnwidth]{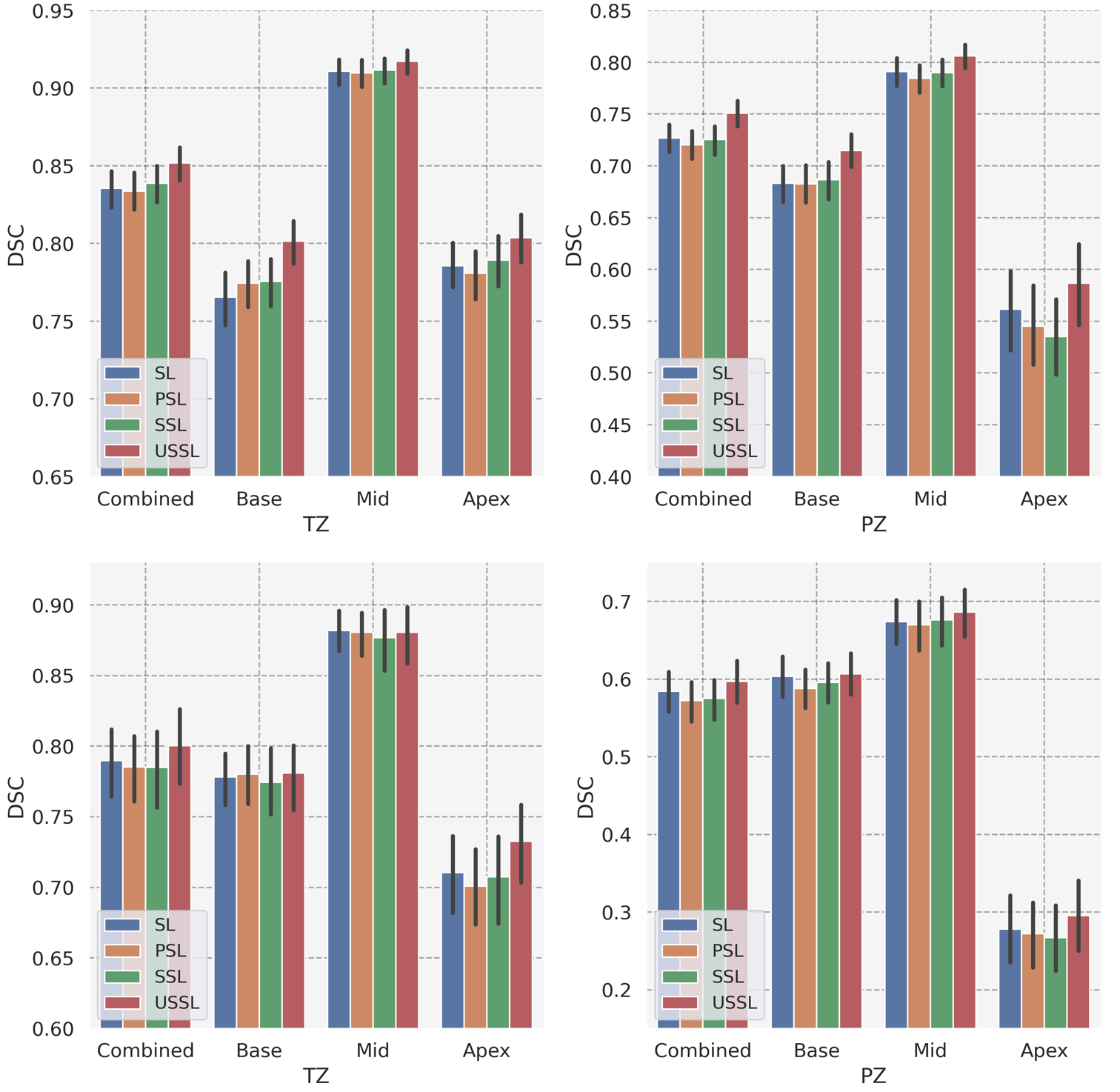}}
\caption{Comparison of Dice Similarity Coefficient(DSC) in different zones and regions using different segmentation methods.\textbf{(\textit{top-row})} ProstateX test set \textbf{(\textit{bottom-row})} the external test set. \textbf{(\textit{DSL})} Attention 3D-UNet Supervised Learning, \textbf{(\textit{SL})} Probabilistic Supervised Learning, \textbf{(\textit{SSL})} Semi-Supervised Learning and \textbf{(\textit{USSL})} Uncertainly-Aware Semi-Supervised Learning }
\label{fig:dsc_all}
\end{figure}

We evaluated the proposed model through both qualitative and quantitative methods. All reported metrics were computed in 3D space over 5-fold cross-validation. For the probabilistic models, we used the mean of 20 executions for inference per image. \ref{fig:visual_results} provides some examples of T2w images, the corresponding manual segmentations, and the segmentations obtained with our proposed model.

\begin{table*}[]
\begin{minipage}{\textwidth}
\centering
\caption{Segmentation DSC accuracy of different methods on the testing datasets}
\renewcommand{\arraystretch}{1.5}
\begin{tabular}{|c|c|cccccccccccc|}
\hline
\multirow{3}{*}{Method}            & \multirow{3}{*}{Dataset} & \multicolumn{12}{c|}{Dice Similarity Score (DSC)}                                                                                                                                                                                                                                                                                         \\ \cline{3-14} 
                                   &                          & \multicolumn{4}{c|}{WG}                                                                                             & \multicolumn{4}{c|}{TZ}                                                                                             & \multicolumn{4}{c|}{PZ}                                                                        \\ \cline{3-14} 
                                   &                          & \multicolumn{1}{c|}{Overall} & \multicolumn{1}{c|}{Base}  & \multicolumn{1}{c|}{Mid}   & \multicolumn{1}{c|}{Apex}  & \multicolumn{1}{c|}{Overall} & \multicolumn{1}{c|}{Base}  & \multicolumn{1}{c|}{Mid}   & \multicolumn{1}{c|}{Apex}  & \multicolumn{1}{c|}{Overall} & \multicolumn{1}{c|}{Base}  & \multicolumn{1}{c|}{Mid}   & Apex  \\ \hline
\multirow{2}{*}{Attention 3D-UNet} & ProstateX                & \multicolumn{1}{c|}{$0.876^\ast$}   & \multicolumn{1}{c|}{0.854} & \multicolumn{1}{c|}{0.925} & \multicolumn{1}{c|}{0.828} & \multicolumn{1}{c|}{$0.835^\ast$}   & \multicolumn{1}{c|}{0.765} & \multicolumn{1}{c|}{0.911} & \multicolumn{1}{c|}{0.786} & \multicolumn{1}{c|}{$0.727^\ast$}   & \multicolumn{1}{c|}{0.683} & \multicolumn{1}{c|}{0.791} & 0.562 \\ \cline{2-14} 
                                   & External                 & \multicolumn{1}{c|}{$0.817^\ast$}   & \multicolumn{1}{c|}{0.829} & \multicolumn{1}{c|}{0.915} & \multicolumn{1}{c|}{0.715} & \multicolumn{1}{c|}{$0.790^\ast$}   & \multicolumn{1}{c|}{0.778} & \multicolumn{1}{c|}{0.882} & \multicolumn{1}{c|}{0.710} & \multicolumn{1}{c|}{$0.584^\ast$}   & \multicolumn{1}{c|}{0.604} & \multicolumn{1}{c|}{0.674} & 0.278 \\ \hline
\multirow{2}{*}{PA-UNet}           & ProstateX                & \multicolumn{1}{c|}{$0.871^\ast$}   & \multicolumn{1}{c|}{0.853} & \multicolumn{1}{c|}{0.921} & \multicolumn{1}{c|}{0.915} & \multicolumn{1}{c|}{$0.834^\ast$}   & \multicolumn{1}{c|}{0.774} & \multicolumn{1}{c|}{0.910} & \multicolumn{1}{c|}{0.781} & \multicolumn{1}{c|}{$0.720^\ast$}   & \multicolumn{1}{c|}{0.683} & \multicolumn{1}{c|}{0.784} & 0.545 \\ \cline{2-14} 
                                   & External                 & \multicolumn{1}{c|}{$0.809^\ast$}   & \multicolumn{1}{c|}{0.817} & \multicolumn{1}{c|}{0.906} & \multicolumn{1}{c|}{0.706} & \multicolumn{1}{c|}{$0.785^\ast$}   & \multicolumn{1}{c|}{0.780} & \multicolumn{1}{c|}{0.881} & \multicolumn{1}{c|}{0.701} & \multicolumn{1}{c|}{$0.573^\ast$}   & \multicolumn{1}{c|}{0.588} & \multicolumn{1}{c|}{0.670} & 0.272 \\ \hline
\multirow{2}{*}{SSL}               & ProstateX                & \multicolumn{1}{c|}{$0.875^\ast$}   & \multicolumn{1}{c|}{0.856} & \multicolumn{1}{c|}{0.923} & \multicolumn{1}{c|}{0.819} & \multicolumn{1}{c|}{$0.839^\ast$}   & \multicolumn{1}{c|}{0.776} & \multicolumn{1}{c|}{0.912} & \multicolumn{1}{c|}{0.789} & \multicolumn{1}{c|}{$0.726^\ast$}   & \multicolumn{1}{c|}{0.687} & \multicolumn{1}{c|}{0.790} & 0.535 \\ \cline{2-14} 
                                   & External                 & \multicolumn{1}{c|}{$0.805^\ast$}   & \multicolumn{1}{c|}{0.816} & \multicolumn{1}{c|}{0.905} & \multicolumn{1}{c|}{0.705} & \multicolumn{1}{c|}{$0.785^\ast$}   & \multicolumn{1}{c|}{0.774} & \multicolumn{1}{c|}{0.877} & \multicolumn{1}{c|}{0.707} & \multicolumn{1}{c|}{$0.575^\ast$}   & \multicolumn{1}{c|}{0.596} & \multicolumn{1}{c|}{0.676} & 0.267 \\ \hline
\multirow{2}{*}{USSL}              & ProstateX                & \multicolumn{1}{c|}{$0.885^\ast$}   & \multicolumn{1}{c|}{0.866} & \multicolumn{1}{c|}{0.928} & \multicolumn{1}{c|}{0.834} & \multicolumn{1}{c|}{$0.852^\ast$}   & \multicolumn{1}{c|}{0.801} & \multicolumn{1}{c|}{0.917} & \multicolumn{1}{c|}{0.804} & \multicolumn{1}{c|}{$0.751^\ast$}   & \multicolumn{1}{c|}{0.715} & \multicolumn{1}{c|}{0.806} & 0.587 \\ \cline{2-14} 
                                   & External                 & \multicolumn{1}{c|}{$0.832^\ast$}   & \multicolumn{1}{c|}{0.825} & \multicolumn{1}{c|}{0.912} & \multicolumn{1}{c|}{0.754} & \multicolumn{1}{c|}{$0.800^\ast$}   & \multicolumn{1}{c|}{0.781} & \multicolumn{1}{c|}{0.881} & \multicolumn{1}{c|}{0.733} & \multicolumn{1}{c|}{$0.597^\ast$}   & \multicolumn{1}{c|}{0.607} & \multicolumn{1}{c|}{0.687} & 0.295 \\ \hline
\end{tabular}

\label{tab:results}
\begin{tablenotes}
  \item[*] $\ast$ statistically significant ($p<0.05$)
\end{tablenotes}
\end{minipage}
\end{table*}

\subsection{Uncertainty estimation} 
We used the estimated uncertainties to rank the predicted segmentations in an unsupervised way. In Fig. \ref{fig:dsc_trend} the accuracy metrics are plotted over the fraction of retained test data. This figure shows model uncertainty was higher for predictions with lower DSC (inverse correlation), and the method can rank the segmentation performance. Methods that are making better estimates of uncertainty show this by improving performance (i.e. DSC) as the portion of retained data decreases by showing steeper slopes in Fig. \ref{fig:dsc_trend}. Table \ref{tab:results} and Fig. \ref{fig:dsc_trend} suggest that the proposed method captures meaningful estimates for uncertainty.

Additionally, we performed an analysis of the relationship between uncertainty estimates and segmentation quality. We found a strong correlation between higher uncertainty values and lower DSC, confirming that the model is effectively using uncertainty to guide the selection of pseudo-labels.

We investigate the dataset shift between in-distribution (ProstateX) and out-distribution (External testset) datasets. Fig. \ref{fig:hist} shows a considerable covariant shift. Despite this shift, our model was able to maintain its performance, suggesting that the USSL approach can handle variations in data distributions.

\subsection{Comparison with other semi-supervised methods}
Table \ref{tab:results} and Fig. \ref{fig:dsc_all} compares our proposed model with various methods, described in section \ref{Comparison with state-of-the-art methods} in terms of segmentation performance measured by DSC (mean $\pm$ std.) for the segmentation of prostate whole-gland and its two zones, TZ and PZ.

Our proposed model consistently outperformed the other methods in terms of DSC across all the prostate zones, indicating the effectiveness of our USSL approach. The improvement was particularly notable in the peripheral zone, where our model achieved a higher DSC compared to other semi-supervised methods. This indicates that our model is particularly well-suited for segmenting more challenging regions.

Furthermore, our proposed model demonstrated superior performance in comparison to fully supervised models. This result highlights the potential of our method to reduce annotation efforts while still achieving state-of-the-art performance.

\section{Discussion}
In this study, we present a semi-supervised approach for prostate zone segmentation that takes uncertainty into consideration and can quantify the predictive uncertainty of the model's segmentation predictions without using ground-truth labels.
Our results reveal that the estimated uncertainty metric obtained using our probabilistic model has an inverse correlation with quantitative metrics such as DSC, allowing us to rank the performance of our predictions and simulate human clinical quality control. 
Importantly, our approach does not require the training of multiple models, as is required in deep ensembling methods.

Our method, referred to as USSL, leverages the availability of unlabeled data to reduce epistemic uncertainty and improve segmentation quality. We show that USSL outperforms both the supervised-only model and standard SSL by utilizing a subset of the unlabeled data. 
Table \ref{tab:results} illustrates the improvement in performance when incorporating 25\% of the unlabeled data in our USSL approach. Overall, our method demonstrates the potential to effectively utilize unlabeled data and improve segmentation quality by incorporating uncertainty estimation. We performed a statistical significance test on the model's performance, and the results show that the USSL model is significantly better than the other models ($p < 0.05$). This indicates that the improvements observed with the USSL model are not due to chance and provide evidence for the effectiveness of our approach.

Our findings indicate that high model uncertainty is often indicative of erroneous predictions, and that this information can be leveraged to improve the performance of the semi-supervised model by selecting a subset of cases with the highest quality pseudo labels.
In \ref{fig:visual_results}, we provide examples of the supervised and USSL results, alongside with the corresponding estimated uncertainty maps. As depicted in the figure, the USSL model produces less uncertainty in ambiguous areas.

Our work demonstrates the feasibility of using uncertainty measures to provide interpretable and informative insights into the quality of deep learning-based predictions for prostate zonal segmentation. We were able to compute meaningful uncertainty measures without the need for additional labels for an explicit uncertain category. 
We applied uncertainty estimation using PUNet for prostate zonal segmentation and found that it was efficient. Running the model 20 times to extract the uncertainty took approximately 3.5 seconds to compute for a single image.

The results presented in this paper have some limitations. All scans used in this study were collected using MRI scanners from a single manufacturer. Although we believe our method should be applicable to other MRI scanners, some of the settings may need further tuning when applied to multi-vendor datasets. Another limitation of this study was the precision of the ground-truth segmentations used to develop and evaluate our models. As we mentioned above, several studies have reported high inter-observer variability for prostate zonal segmentation in T2w images \cite{becker2019variability,montagne2021challenge}. To address this issue, it would be beneficial to obtain zonal segmentation labels that reflect the consensus of multiple experts for a large-scale prostate MRI dataset.

Despite the challenges and limitations inherent in uncertainty-based semi-supervised learning approaches, the demonstrated performance of our proposed method, its requirement for only a small portion of labeled data, and its relative simplicity suggest that it is a promising approach for use in prostate zonal segmentation.
In the future, we aim to apply our framework to other semi-supervised medical image segmentation tasks. Our results contribute to the growing evidence that the development of deep learning applications often requires large training datasets, and that semi-supervised learning can be particularly beneficial when the ratio of annotated to unannotated images is small, as is commonly the case in medical imaging.

\section{Conclusion}
We present a novel uncertainty-aware semi-supervised learning (USSL) method for the segmentation of prostate zones in T2-weighted MRI images. We demonstrate that the segmentations produced by USSL exhibit superior quality when compared to the same model employing standard SSL methods.
Moreover, we modeled predictive segmentation uncertainty using a probabilistic model which can generate a set of plausible segmentations. Furthermore, we explore the predictive uncertainty to improve the quality of our segmentations by guiding a semi-supervised model. The results of our experiments show that the proposed method performs better segmentation compared to the supervised and semi-supervised methods. Our findings emphasize the importance of incorporating uncertainty estimation in deep learning-based medical image segmentation tasks, particularly in cases where labeled data is scarce.


\bibliographystyle{IEEEtran}
\bibliography{main}

\end{document}